\documentclass[12pt]{article}
\usepackage{epsfig,amssymb,amsmath,psfrag}
\usepackage{cite}

\font\cmss=cmss10 at 11pt \font\cmsss=cmss8 at 8pt

\def\mininbar{\vrule height.75ex width.3pt depth0pt}
\def\cc{\relax\,\hbox{$\mininbar\kern-.2em{\hbox{\rm\tiny C}}$}}
\def\Z{\relax\ifmmode\mathchoice
{\hbox{\cmss Z\kern-.4em Z}}{\hbox{\cmss Z\kern-.4em Z}}
{\lower.4pt\hbox{\cmsss Z\kern-.4em Z}}
{\lower1.2pt\hbox{\cmsss Z\kern-.4em Z}}\else{\cmss Z\kern-.4em Z}\fi}
\def\one{1\kern -3pt \mathrm{l}}

\def\ph{ \phantom{-}1 }
\def\T{t} 
\def\M{M}

\def\f{\tilde{f}}
\def\tn{\tilde{n}}
\def\tbeta{\tilde{\beta}}
\def\tgamma{\tilde{\gamma}}
\def\eps{\epsilon}
\def\Zero{{(0)}}
\def\One{{(1)}}

\def\cN {  {\cal N}  }
\def\cA {  {\cal A}_5^\One  }
\def\cM {  {\cal M}_5^\One  }
\def\cO {  {\cal O}  }

\def\la {\lambda}

\def\eg{{e.g.}}

\def\eqn#1{eq.~(\ref{#1})} \def\Eqn#1{Equation~(\ref{#1})}
\def\eqns#1#2{eqs.~(\ref{#1}) and~(\ref{#2})}

\def \be  {\begin{equation}}
\def \ee  {\end{equation}}
\def \ba  {\begin{eqnarray}}
\def \ea  {\end{eqnarray}}
\def \Tr {\mathop{\rm Tr}\nolimits}

\newcommand{\nn}{\nonumber}

\begin{document}

\begin{flushright}
BOW-PH-151\\
BRX-TH-639
\end{flushright}
\vspace{3mm}

\begin{center}
{\Large\bf\sf
One-loop SYM-supergravity relation 
for five-point amplitudes}

\vskip 10mm
Stephen G. Naculich\footnote{
Research supported in part by the NSF under grant PHY-0756518}$^{,a}$
and 
Howard J. Schnitzer\footnote{
Research supported in part by the DOE under grant DE--FG02--92ER40706
}$^{,b}$
\end{center}

\begin{center}
$^{a}${\em Department of Physics\\
Bowdoin College, Brunswick, ME 04011, USA}

\vspace{5mm}

$^{b}${\em Theoretical Physics Group\\
Martin Fisher School of Physics\\
Brandeis University, Waltham, MA 02454, USA}

\vspace{5mm}
{\tt naculich@bowdoin.edu, schnitzr@brandeis.edu}
\end{center}

\vskip 2mm

\begin{abstract}

We derive a linear relation between the
one-loop five-point amplitude of $\cN=8$ supergravity 
and the
one-loop five-point subleading-color amplitudes
of $\cN=4$ supersymmetric Yang-Mills theory.
\end{abstract}
\vfil\break

\section{Introduction}
\setcounter{equation}{0}

The recent explosion of interest in 
$\cN=4$ supersymmetric Yang-Mills  (SYM)
and $\cN=8$ supergravity amplitudes 
is due to the discovery of a host of 
symmetries and structures that are completely hidden 
in the traditional Feynman diagram 
approach to computing amplitudes.
Perhaps even more surprising are the relations 
that have begun to emerge between perturbative gauge theory 
and gravity amplitudes.

Tree-level relations between gauge and graviton amplitudes 
were originally discovered
in the field-theory limit of the string-theoretic 
KLT relations \cite{Kawai:1985xq}.
One-loop relations 
between $\cN=4$ SYM and $\cN=8$ supergravity
amplitudes were also first obtained using string theory
\cite{Green:1982sw},
while unitarity methods have been used to 
derive relations at higher loops
\cite{Bern:1998ug,Bern:1998sv,Bern:2008pv,Bern:2010tq}.
While much of the recent spectacular progress has
involved planar (large-$N$) SYM amplitudes,
the connection between SYM and supergravity amplitudes
intimately involves the non-planar (subleading-in-$1/N$) 
contributions to SYM amplitudes.

The connection between gauge and gravity amplitudes 
has recently been significantly tightened
through the discovery of a new color-kinematic duality 
of gauge theory amplitudes by Bern, Carrasco, and Johansson
\cite{Bern:2008qj,Bern:2010ue}.  
It was conjectured \cite{Bern:2008qj} and proven 
\cite{BjerrumBohr:2009rd,Stieberger:2009hq,Feng:2010my,Chen:2011jx}
that tree-level $n$-gluon amplitudes may be written as a sum
over diagrams built from cubic vertices 
\be
{\cal A} =  \sum_i {n_i c_i \over  d_i}
\label{bcj}
\ee
in which the kinematic numerators $n_i$ obey precisely
the same set of algebraic relations observed by the color
factors $c_i$.  
The denominators $d_i$ are products of the inverse propagators 
corresponding to each cubic diagram.
Given a tree-level $n$-gluon amplitude in a form 
that respects color-kinematic duality, 
it was further conjectured \cite{Bern:2008qj}  and proven \cite{Bern:2010yg}
that $n$-graviton amplitudes can be expressed as a sum over the same diagrams
\be
{\cal M} =  \sum_i {\tn_i n_i \over d_i}
\ee
with the color factors replaced by a second copy $\tn_i$ of 
the numerator factors.
At the loop level,
color-kinematic duality was conjectured \cite{Bern:2008qj,Bern:2010ue} 
to hold for the numerator factors appearing in the integrands
of gauge-theory loop diagrams built from cubic vertices,
and this was verified through three loops 
for the $\cN=4$ SYM four-point amplitude \cite{Bern:2010ue,Carrasco:2011hw}.
Loop-level gravity amplitudes can then
be obtained from the same diagrams
using a double copy of the numerator factors.
Related work on color-kinematic duality appears in refs.~\cite{
Sondergaard:2009za,
Broedel:2011pd}, 
and some recent reviews are refs.~\cite{Carrasco:2011hw,Sondergaard:2011iv}.

Recently, Carrasco and Johansson \cite{Carrasco:2011mn} 
demonstrated that the five-point amplitude 
can be written in a form in which the numerators
of the integrands respect color-kinematic duality
at one, two\cite{Carrasco:2011mn}, and three \cite{Carrasco} loops.
The one-loop $\cN=4$ SYM five-point amplitude is \cite{Carrasco:2011mn}
\be
\cA
= 
 i g^5 \sum_{S_5} \left( 
\frac{1}{10} \beta_{12345} I^{(P)} (12345) C^{(P)}_{12345}  
+
\frac{1}{4} \gamma_{12}  I^{(B)} (12;345) C^{(B)}_{12;345}  \right)
\label{cjintro}
\ee
where $I^{(P)}$ and $I^{(B)}$ 
are the scalar pentagon and box integrals 
defined below in \eqns{Ipentagon}{Ibox},
$C^{(P)}$ and $C^{(B)}$ 
are the corresponding color factors
defined in \eqns{pentagon}{boxplusline},
and\footnote{A representation of the 
tree-level five-point amplitude in terms of 
$\beta_{12345}$ and $\gamma_{12}$ was presented in
ref.~\cite{Broedel:2011pd}. }
\be
\beta_{12345} =
\delta^{(8)}(Q) {[12][23][34][45][51] \over 4 \, \varepsilon(1,2,3,4)},
\qquad\qquad
\gamma_{12} =
\delta^{(8)}(Q) {[12]^2 [34][45][35] \over 4 \, \varepsilon(1,2,3,4)}
\label{betaandgamma}
\ee
where $[ij]$ are the usual helicity spinor products,
$\varepsilon(1,2,3,4)\equiv 
\varepsilon_{\mu\nu\rho\sigma}k_1^\mu k_2^\nu k_3^\rho k_4^\sigma$,
and $ \delta^{(8)}(Q) $ is the 
manifestly-supersymmetric delta function of 
the Grassmann-valued supermomentum. 
The BCJ conjecture \cite{Bern:2010ue}
then suggests that the corresponding one-loop 
$\cN=8$ supergravity five-point amplitude is \cite{Carrasco:2011mn}
\be
\cM = 
- \left(\kappa \over 2\right)^5 \sum_{S_5} 
\left( \frac{1}{10} \tbeta_{12345} \beta_{12345} I^{(P)} (12345) 
+ \frac{1}{4} \tgamma_{12}  \gamma_{12}  I^{(B)} (12;345) \right)
\label{sgvintro}
\ee
where $\tbeta_{12345} $ and $\tgamma_{12} $
are numerator factors of a second copy of $\cN=4$ SYM,
and it was verified \cite{Carrasco:2011mn,Bern:2011rj}
that \eqn{sgvintro} agrees with the known expression in ref.~\cite{Bern:1998sv}.

Loop amplitudes of $\cN=4$ SYM theory possess IR divergences
which can be regulated through dimensional regularization
in $D = 4- 2 \eps$ dimensions with $\eps < 0$.
The leading IR divergence of one-loop $n$-point amplitudes,
such as \eqn{cjintro},
goes as $1/\eps^2$ \cite{Bern:1994zx}.
In a $1/N$ expansion of the SYM amplitude, however, 
the coefficients of the subleading-color amplitudes 
have IR divergences that are less 
severe than the leading-color 
(planar) amplitudes\cite{Naculich:2008ys,Naculich:2009cv,Nastase:2011mx}.
At one loop, for example, the IR divergence of the 
double-trace amplitude (which is subleading in $1/N$) goes as $1/\eps$,
as ascertained for $n$-point functions in ref.~\cite{Nastase:2011mx}.
This matches the leading IR divergence
of one-loop $\cN=8$ supergravity amplitudes,
which also go as $1/\eps$ \cite{Dunbar:1995ed},
suggesting the possibility of linear relations
between supergravity amplitudes
and subleading-color SYM amplitudes.
Such relations were found 
at one and two loops for four-point amplitudes
in ref.~\cite{Naculich:2008ys}.
In this paper, we propose and prove the relation
\be
\cM = {-1\over 20 i g^5} \left(\kappa \over 2\right)^5
\sum_{S_5} \tbeta_{12345}  A_{5;3} (12;345)
\ee
between the one-loop $\cN=8$ supergravity five-point amplitude $\cM$
and the subleading-color SYM amplitude $A_{5;3} (12;345)$,  
defined 
as the coefficient of the double-trace term in
the trace expansion of 
the one-loop $\cN=4$ SYM five-point amplitude $\cA$
(cf. \eqn{oneloopfivepointtrace}). 
This relation is distinct from one proposed in 
ref.~\cite{Bern:2011rj} between the one-loop supergravity amplitude 
and the leading-color SYM amplitudes $A_{5;1} (12345)$,  
as we discuss at the end of sec.~\ref{secsgvsym}.

The relation between loop-level gauge theory and 
supergravity amplitudes proposed in 
ref.~\cite{Bern:2010ue} 
is a diagram-by-diagram map between the {\it integrands}.
In contrast,
the various relations between subleading-color SYM and supergravity amplitudes 
found for one- and two-loop four-point amplitudes in 
ref.~\cite{Naculich:2008ys},
for one-loop four- and five-point amplitudes in ref.~\cite{Bern:2011rj},
and for one-loop five-point amplitudes in the present paper,
are between the {\it integrated} amplitudes.
We observe that in all of these cases, 
the kinematic numerators are independent 
of the loop momenta and therefore can be factored out, 
leaving the same set of scalar integrals
contributing to both the SYM and supergravity amplitudes.
Finding relations between the 
integrated SYM  and supergravity amplitudes when the numerator factors 
are dependent on the loop momenta will be more challenging.

This paper is organized as follows.
In section 2, we review the color and trace
bases for representing SYM amplitudes. 
In section 3, we decompose the Carrasco-Johannson 
representation of the one-loop five-point amplitude 
into the trace basis.
In section 4, we present and prove a linear relation
between the one-loop five-point supergravity amplitude
and the one-loop subleading-color five-point SYM amplitudes.
Section 5 contains our conclusions,
and in an appendix we review the leading IR divergence 
of the one-loop subleading-color five-point SYM amplitude.

\section{The one-loop $\cN=4$ SYM five-point amplitude} 
\setcounter{equation}{0}

In this section, we review how the one-loop five-point amplitude
of $\cN=4$ SYM theory
may be expressed in both the color basis  and the trace basis,
and how the coefficients in each of these bases
are related to one another.

\begin{figure}[t]
\centerline{
{\epsfxsize4cm  \epsfbox{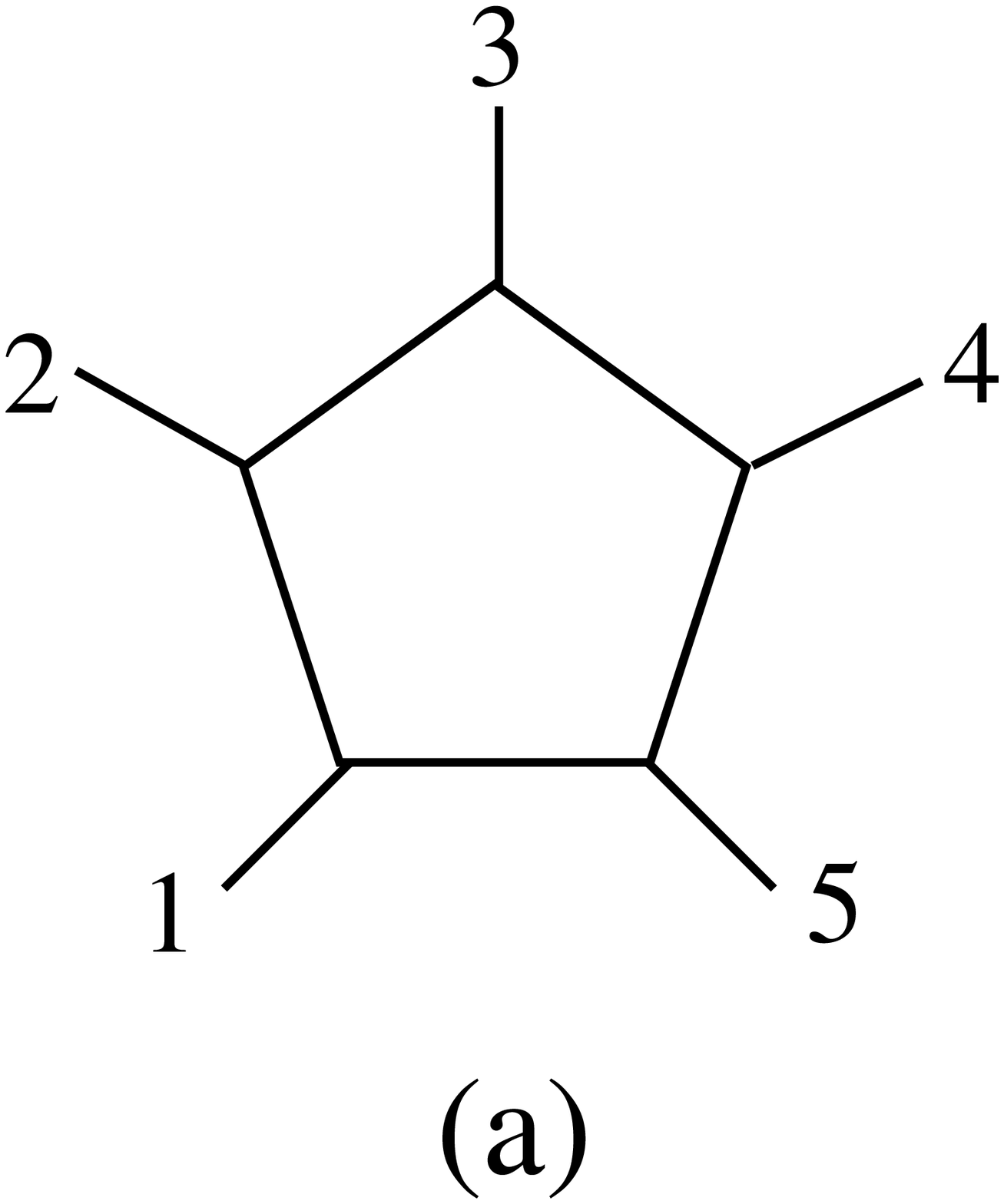}}
\hskip2cm
{\epsfxsize4cm  \epsfbox{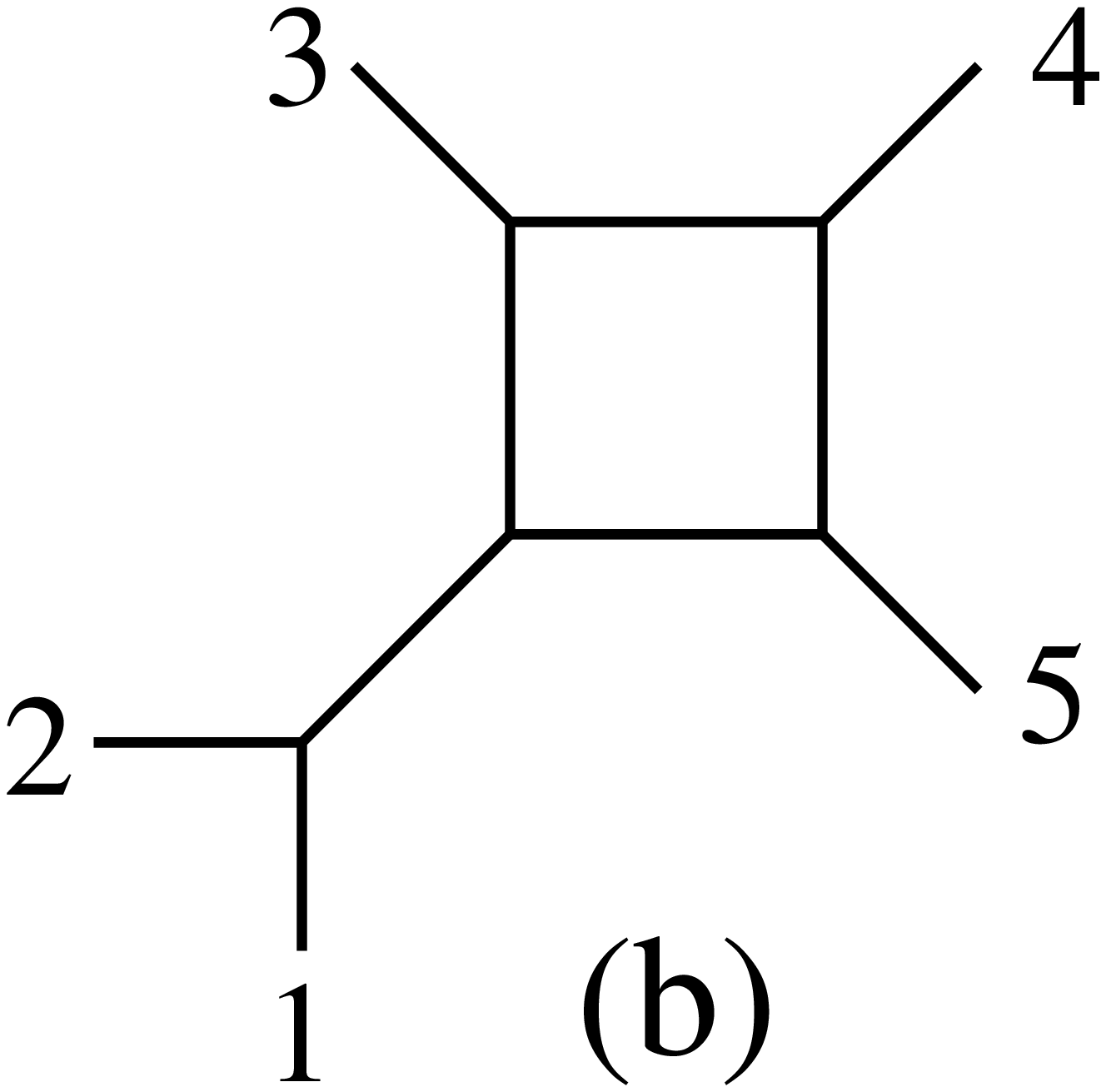}}
}
\caption{One-loop five-point diagrams} 
\label{diagrams}
\end{figure}

A complete color basis for one-loop five-point amplitudes 
is spanned by the pentagon color diagram in fig.~\ref{diagrams}(a) 
\be
C_{12345}^{(P)} =
\f^{g a_1 b} \f^{b a_2 c} \f^{c a_3 d} \f^{d a_4 e} \f^{e a_5 g} 
\label{pentagon}
\ee
and its permutations (\eg, see ref.~\cite{DelDuca:1999rs}),
where $\f^{abc} = i \sqrt{2} f^{abc} $
are the structure constants for the SU$(N)$ gauge group. 
The one-loop five-point amplitude can then be expressed as
\be
\cA = \sum_{S_5}  {1 \over 10} a(12345) \,C_{12345}^{(P)} 
\label{fivepointamp}
\ee
where  $a(12345)$ depends on the momenta and helicities 
of the external states, 
and the sum is over all permutations
of the momentum, helicity, and color index $a$ of the
external states. 
The color factor
$C_{12345}^{(P)}$
is manifestly 
symmetric under cyclic permutations of $12345$ and 
antisymmetric under $12345 \to 54321$. 
Bose symmetry implies that
$a(12345)$ is therefore also symmetric under cyclic permutations
and antisymmetric under reversal of indices, 
so we can rewrite the amplitude as a sum over twelve terms 
\be
\cA = \sum_{S_5/\Z_5 \times \Z_2}  a(12345) \,C_{12345}^{(P)}
= \sum_{i=1}^{12} a_i c_i 
\label{oneloopfivepointcolor}
\ee
where we choose an explicit basis
\be
c_i =  \{ 
C^{(P)}_{12345}, C^{(P)}_{14325} , C^{(P)}_{13425} , 
C^{(P)}_{12435} , C^{(P)}_{14235} , C^{(P)}_{13245} ,
C^{(P)}_{12543} , C^{(P)}_{14523} , C^{(P)}_{13524} , 
C^{(P)}_{12534} , C^{(P)}_{14532} , C^{(P)}_{13542} 
\}
\label{colorbasis}
\ee
for later convenience. 

Alternatively, the one-loop five-point amplitude may be expressed 
in terms of a trace basis \cite{Bern:1990ux}
\ba
\cA 
&=&
\sum_{S_5/\Z_5 \times \Z_2} 
A_{5;1} (12345) \, N \left[\Tr(12345) - \Tr(54321)\right]
\nn\\
&&
+ \sum_{S_5/\Z_2 \times S_3} 
A_{5;3} (12;345) \Tr(12) \left[\Tr(345) - \Tr(543)\right]
\label{oneloopfivepointtrace}
\ea
where $\Tr(12345) \equiv \Tr( T^{a_1} T^{a_2} T^{a_3} T^{a_4} T^{a_5})$
and the matrices $T^a$  are the generators in the defining 
representation of SU$(N)$, normalized according to 
$\Tr (T^a T^b) = \delta^{ab}$.

There are twelve independent single-trace coefficients 
$A_{5;1} (12345)$,
which are 
symmetric under cyclic permutations and 
antisymmetric under reversal 
of indices,
and ten independent double-trace coefficients
$A_{5;3} (12;345)$, 
which are symmetric under exchange of the first two indices,
and completely antisymmetric under permutations of the last three indices. 
We therefore write
\be
\cA = \sum_{\la=1}^{22}  A_\la \T_\la
\ee
where the basis $\{ \T_\la \}$ consists of the following single-trace terms
\ba
\T_1 	&=&  N\left[\Tr (12345) - \Tr(15432)\right] \qquad\qquad
\T_7 	  =  N\left[\Tr (12543) - \Tr(13452)\right] \nonumber\\
\T_2 	&=&  N\left[\Tr (14325) - \Tr(15234)\right] \qquad\qquad
\T_8 	  =  N\left[\Tr (14523) - \Tr(13254)\right] \nonumber\\
\T_3 	&=&  N\left[\Tr (13425) - \Tr(15243)\right] \qquad\qquad
\T_9 	  =  N\left[\Tr (13524) - \Tr(14253)\right] \nonumber\\
\T_4 	&=&  N\left[\Tr (12435) - \Tr(15342)\right] \qquad\qquad
\T_{10}   =  N\left[\Tr (12534) - \Tr(14352)\right] \nonumber\\
\T_5 	&=&  N\left[\Tr (14235) - \Tr(15324)\right] \qquad\qquad
\T_{11}   =  N\left[\Tr (14532) - \Tr(12354)\right] \nonumber\\
\T_6 	&=&  N\left[\Tr (13245) - \Tr(15423)\right] \qquad\qquad
\T_{12}   =  N\left[\Tr (13542) - \Tr(12453)\right] 
\label{singletrace}
\ea
and the following double-trace terms 
\ba
\T_{13} &=&  \Tr (12) \left[ \Tr(345) - \Tr(543) \right]  \qquad\qquad
\T_{18}   =  \Tr (13) \left[ \Tr(245) - \Tr(542) \right]  \nonumber\\
\T_{14} &=&  \Tr (23) \left[ \Tr(451) - \Tr(154) \right]  \qquad\qquad
\T_{19}   =  \Tr (24) \left[ \Tr(351) - \Tr(153) \right]  \nonumber\\
\T_{15} &=&  \Tr (34) \left[ \Tr(512) - \Tr(215) \right]  \qquad\qquad
\T_{20}   =  \Tr (35) \left[ \Tr(412) - \Tr(214) \right]  \nonumber\\
\T_{16} &=&  \Tr (45) \left[ \Tr(123) - \Tr(321) \right]  \qquad\qquad
\T_{21}   =  \Tr (41) \left[ \Tr(523) - \Tr(325) \right]  \nonumber\\
\T_{17} &=&  \Tr (51) \left[ \Tr(234) - \Tr(432) \right]  \qquad\qquad
\T_{22}   =  \Tr (52) \left[ \Tr(134) - \Tr(431) \right]   \, .
\label{doubletrace}
\ea
The color (\ref{colorbasis}) and trace bases (\ref{singletrace}) and
(\ref{doubletrace}) are related by
\be
c_i = \sum_{\la=1}^{22}  \M_{i\la} \T_\la \, .
\label{trans}
\ee
Using $\f^{abc} = \Tr( [T^a, T^b]  T^c )$ to evaluate \eqn{pentagon} 
yields 
\be
c_1 = \T_1 + \sum_{\la=13}^{22} \T_\la
\ee
which gives the first row  $\M_{1\la}$ of the transformation matrix.
The other rows may be obtained through permutations of the 
external state indices
\be
\M_{i\la} 
= \left(
   \begin{array}{cccccccccccccccccccccc}
    1 & 0 & 0 & 0 & 0 & 0 & 0 & 0 & 0 & 0 & 0 & 0~
 & \ph & \ph & \ph & \ph & \ph & \ph & \ph & \ph & \ph & \ph \\
    0 & 1 & 0 & 0 & 0 & 0 & 0 & 0 & 0 & 0 & 0 & 0~
 & -1 & \ph & \ph & -1 & -1 & -1 &\ph & -1 & -1 & -1 \\
    0 & 0 & 1 & 0 & 0 & 0 & 0 & 0 & 0 & 0 & 0 & 0~
 &\ph &\ph &\ph & -1 &\ph & -1 & \ph & -1 & -1 &\ph \\
    0 & 0 & 0 & 1 & 0 & 0 & 0 & 0 & 0 & 0 & 0 & 0~
 & -1 &\ph &\ph &\ph & -1 &\ph & \ph &\ph &\ph & -1 \\
    0 & 0 & 0 & 0 & 1 & 0 & 0 & 0 & 0 & 0 & 0 & 0~
 & -1 &\ph &\ph &\ph &\ph & -1 & \ph & -1 &\ph & -1 \\
    0 & 0 & 0 & 0 & 0 & 1 & 0 & 0 & 0 & 0 & 0 & 0~
 &\ph &\ph &\ph & -1 & -1 &\ph & \ph &\ph & -1 &\ph \\
    0 & 0 & 0 & 0 & 0 & 0 & 1 & 0 & 0 & 0 & 0 & 0~
 & -1 & -1 &\ph &\ph & -1 & -1 & -1 &\ph & -1 & -1 \\
    0 & 0 & 0 & 0 & 0 & 0 & 0 & 1 & 0 & 0 & 0 & 0~
 &\ph &\ph & -1 &\ph &\ph &\ph & -1 & -1 &\ph & -1 \\
    0 & 0 & 0 & 0 & 0 & 0 & 0 & 0 & 1 & 0 & 0 & 0~
 & -1 & -1 & -1 & -1 & -1 &\ph &\ph &\ph &\ph &\ph \\
    0 & 0 & 0 & 0 & 0 & 0 & 0 & 0 & 0 & 1 & 0 & 0~
 &\ph & -1 &\ph &\ph &\ph & -1 & -1 &\ph & -1 &\ph \\
    0 & 0 & 0 & 0 & 0 & 0 & 0 & 0 & 0 & 0 & 1 & 0~
 &\ph &\ph & -1 & -1 & -1 &\ph & -1 & -1 & -1 & -1 \\
    0 & 0 & 0 & 0 & 0 & 0 & 0 & 0 & 0 & 0 & 0 & 1~ 
& -1 & -1 & -1 & -1 &\ph & -1 &\ph & -1 &\ph &\ph
   \end{array}
\right)
\label{transmatrix}
\ee
\Eqn{trans} implies that the 
coefficients of the color and trace bases are related by
\be
A_\la = \sum_{i=1}^{12} a_i \M_{i\la}
\label{relatecoeff}
\ee
hence
\be
A_i = a_i , \qquad {i=1, \cdots, 12}
\label{singletracecoeff}
\ee
i.e., the coefficients $a(12345)$ in the color 
basis (\ref{oneloopfivepointcolor})
are precisely equal to $A_{5;1} (12345)$, 
the coefficients of the single-trace terms in \eqn{oneloopfivepointtrace}.
On the other hand, 
these planar amplitudes are well-known to be given by the sum of five 
1m scalar box integrals \cite{Bern:1994zx}.
The leading IR divergence of these planar amplitudes therefore 
goes as $1/\eps^2$.

Equations (\ref{transmatrix}) and (\ref{relatecoeff})
also give the coefficients of 
the double-trace terms; \eg,
\be
A_{13} = 
 a_{1} -a_{2} +a_{3} -a_{4} 
-a_{5} +a_{6} -a_{7} +a_{8} 
-a_{9} +a_{10} +a_{11} -a_{12}  \, .
\label{doubletracecoeff}
\ee
Comparing \eqns{singletracecoeff}{doubletracecoeff} 
we see that each of the double-trace amplitudes can
be written as a linear combination of the single-trace amplitudes,
a result long known \cite{Bern:1990ux,Bern:1994zx}.
An equivalent way to derive this result is to observe that
the rank-twelve matrix (\ref{transmatrix})
possesses ten null eigenvectors $R_{\la j}$:
\be
\sum_{\la = 1}^{22}   \M_{i\la} R_{\la j} = 0, \qquad  j = 1, \cdots 10 \,.
\ee
Given that $M_{i \la}$ has the form 
$\begin{pmatrix} \one_{12 \times 12}  & m \end{pmatrix}$,
one sees that 
$R = \begin{pmatrix} -m \\ \one_{10 \times 10} \end{pmatrix}$.
The existence of these null eigenvectors, together with \eqn{relatecoeff},
implies ten relations among the color-ordered amplitudes
\be
\sum_{\la=1}^{22}  A_\la R_{\la j} = 0, \qquad  j = 1, \cdots 10 \, .
\ee
For example, the  $j=1$ relation is
\be
A_{13} = 
A_{1} -A_{2} +A_{3} -A_{4} 
-A_{5} +A_{6} -A_{7} +A_{8} -A_{9} +A_{10} +A_{11} -A_{12} 
\label{Athirteen}
\ee
that is,
\ba
 A_{5;3} (12;345) 
&=&
+ A_{5;1} (12345) 
-A_{5;1} (14325) 
+A_{5;1} (13425) 
-A_{5;1} (12435) 
\nn\\ &&
-A_{5;1} (14235) 
+A_{5;1} (13245) 
-A_{5;1} (12543) 
+A_{5;1} (14523) 
\nn \\ &&
-A_{5;1} (13524) 
+A_{5;1} (12534) 
+A_{5;1} (14532) 
-A_{5;1} (13542)  \, .
\label{singledouble}
\ea
Using the symmetries of the single-trace amplitudes, this can be written 
\ba
 A_{5;3} (12;345) 
&=&
+A_{5;1} (12345) 
+A_{5;1} (23415) 
+A_{5;1} (13425) 
+A_{5;1} (34215) 
\nn\\
&&
+A_{5;1} (32415) 
+A_{5;1} (13245) 
+A_{5;1} (21345) 
+A_{5;1} (23145) 
\nn\\
&&
+A_{5;1} (31425) 
+A_{5;1} (34125) 
+A_{5;1} (32145) 
+A_{5;1} (31245) 
\ea
precisely the set of cyclically-ordered permutations
in eq.~(7.3) of ref.~\cite{Bern:1994zx}.

Although the planar amplitudes $A_{5;1} (12345)$ 
have a leading $1/\eps^2$ IR divergence,
\eqn{singledouble} implies that the leading IR divergence 
of the subleading-color amplitude $A_{5;3} (12;345)$ 
is only $1/\eps$,
as shown in ref.~\cite{Nastase:2011mx}.
We rederive this result in an appendix of this paper.

\section{The CJ representation of the five-point amplitude}
\setcounter{equation}{0}

Carrasco and Johansson recently derived expressions for 
one- and two-loop $\cN=4$ SYM five-point amplitudes 
that manifest color-kinematic duality \cite{Carrasco:2011mn}.
At one-loop their ansatz takes the form \cite{Bern:2010ue}
\be
\cA
= i g^5 \sum_{i} \int {d^D p \over (2\pi)^D} 
{1 \over S_i} {N_i C_i \over \prod_m l_{i_m}^2 }
\label{ansatz}
\ee
where the sum is over all cubic one-loop five-point diagrams,
including relabelings of the external lines,
with symmetry factors $S_i$.
The $l_{i_m}$ are the momenta 
flowing through each of the internal legs of the diagram, 
which can depend on the external momenta $k_j$ and the loop momentum $p$.
The numerator factors $N_i$ are as-yet-unspecified functions
of momenta and helicity. 
The color factors $C_i$ include the pentagon color diagram 
previously specified in \eqn{pentagon},
but to obtain a representation of the amplitude that
satisfies color-kinematic duality,
one must also include the box-plus-line color diagram in
fig.~\ref{diagrams}(b) 
\be
C^{(B)}_{12;345}
=
\f^{a_1 a_2 b} \f^{b  c g } \f^{c a_3 d} \f^{d a_4 e} \f^{e a_5 g} 
\label{boxplusline}
\ee
which is manifestly antisymmetric under $1 \leftrightarrow 2$ 
and symmetric under $3 \leftrightarrow 5$.
The box color diagrams are not independent 
of the pentagon color diagrams;  
the Jacobi identity implies
\be
C^{(B)}_{12;345} = C^{(P)}_{12345} + C^{(P)}_{12543}  \,.
\label{pentagonboxrelation}
\ee
Other color diagrams containing triangles and bubbles need not
be included since the corresponding numerator factors $N_i$ 
vanish for $\cN=4$ SYM theory. 

In a beautiful analysis employing supersymmmetry, 
generalized unitarity, and color-kinematic duality of 
the numerator factors $N^{(P)}$ and $N^{(B)}$,
Carrasco and Johannson showed that the numerator factors 
at one loop are independent of the loop momentum,\footnote{For 
the two-loop five-point amplitude, the numerator factors do 
include dependence on the loop momenta.}
and so \eqn{ansatz} can be written as
\be
\cA
= 
 i g^5 \sum_{S_5} \left( 
\frac{1}{10} \beta_{12345} I^{(P)} (12345) C^{(P)}_{12345}  
+
\frac{1}{4} \gamma_{12;345}  I^{(B)} (12;345) C^{(B)}_{12;345}  \right)
\label{cjfirst}
\ee
where $I^{(P)}$ and $I^{(B)}$ are the scalar integrals 
shown in fig.~\ref{diagrams}
\ba
I^{(\rm P)}(12345)
&=&
\int \frac{d^Dp}{(2\pi)^D} 
\frac{1}{p^2(p+k_1)^2(p+k_1+k_2)^2(p-k_4-k_5)^2(p-k_5)^2}\,, 
\label{Ipentagon} \\
I^{(\rm B)}(12345)
&=&
\frac{1}{(k_1+k_2)^2}
\int 
\frac{d^Dp}{(2\pi)^D} 
\frac{1}{p^2(p+k_1+k_2)^2(p-k_4-k_5)^2(p-k_5)^2}\,,
\label{Ibox}
\ea
and $\beta_{12345}$ and $\gamma_{12;345}$ are functions of external momenta 
and helicities given by \eqn{betaandgamma},
although we will not need their explicit forms in what follows.

One may see from \eqns{Ipentagon}{Ibox} that
$I^{(P)}(12345)$ is invariant under cyclic permutations and reversals
of $12345$, 
and $I^{(B)}(12;345)$ is invariant under $1 \leftrightarrow 2$ and $3 \leftrightarrow 5$. 
Hence\footnote{
As we will see below, $\beta_{12345}$ and $\gamma_{12;345}$
possess further symmetries as well.} 
$\beta_{12345}$ is symmetric under cyclic permutations
and antisymmetric under reversal of indices,
and $\gamma_{12;345}$ is antisymmetric under $1 \leftrightarrow 2$
and symmetric under $3 \leftrightarrow 5$.
On the basis of these symmetries, we rewrite \eqn{cjfirst} as 
\be
\cA
= 
i g^5
\left(
\sum_{S_5/\Z_5 \times \Z_2} 
\beta_{12345} I^{(P)} (12345) C^{(P)}_{12345}  
+
\sum_{S_5/\Z_2 \times \Z_2} 
\gamma_{12;345}  I^{(B)} (12;345) C^{(B)}_{12;345} 
\right) \,.
\label{cjsecond}
\ee
We now recast the Carrasco-Johansson one-loop amplitude 
into the trace basis. 
First we substitute \eqn{pentagonboxrelation} into \eqn{cjfirst}
to obtain
\ba
\cA
&=& 
 i g^5 
\sum_{S_5} \left( 
\frac{1}{10} \beta_{12345} I^{(P)} (12345) 
+
\frac{1}{2} \gamma_{12;345}  I^{(B)} (12;345) \right)
C^{(P)}_{12345}  
\\
&=& 
 i g^5 \sum_{S_5 / \Z_5 \times \Z_2}
\bigg(
\beta_{12345} I^{(P)} (12345) 
+
\Big[ 
\gamma_{12;345}  I^{(B)} (12;345) + 
\gamma_{23;451}  I^{(B)} (23;451) + 
\nn\\
&&\hskip2cm
 +
\gamma_{34;512}  I^{(B)} (34;512) + 
\gamma_{45;123}  I^{(B)} (45;123) +
\gamma_{51;234}  I^{(B)} (51;234)  \Big]
 \bigg)
C^{(P)}_{12345}   \,.
\nn
\ea
By comparing with \eqns{oneloopfivepointcolor}{singletracecoeff},
one sees that 
the one-loop planar amplitude 
is given by 
\ba
A_{5;1} (12345)
 &=&
i g^5 
\bigg(
\beta_{12345} I^{(P)} (12345) 
+
\Big[ 
\gamma_{12;345}  I^{(B)} (12;345) + 
\gamma_{23;451}  I^{(B)} (23;451) + 
\hskip2cm
\nn\\
&&\hskip5mm
 +
\gamma_{34;512}  I^{(B)} (34;512) + 
\gamma_{45;123}  I^{(B)} (45;123) +
\gamma_{51;234}  I^{(B)} (51;234)  \Big]
 \bigg) \,.
\label{comparing}
\ea
The double-trace color-ordered amplitude follows
from \eqn{singledouble}:
\ba
A_{5;3}(12;345)
&=&
i g^5 \bigg(
+\beta_{12345} I^{(P)} (12345) 
-\beta_{14325} I^{(P)}(14325) 
+\beta_{13425} I^{(P)}(13425)  \hskip2.5cm
\nn\\
&&
\hskip1cm
-\beta_{12435} I^{(P)}(12435) 
-\beta_{14235} I^{(P)}(14235) 
+\beta_{13245} I^{(P)}(13245) 
\nn\\
&&
\hskip1cm
-\beta_{12543} I^{(P)}(12543) 
+\beta_{14523} I^{(P)}(14523) 
-\beta_{13524} I^{(P)}(13524) 
\nn\\
&&
\hskip1cm
+\beta_{12534} I^{(P)}(12534) 
+\beta_{14532} I^{(P)}(14532) 
-\beta_{13542} I^{(P)}(13542) 
\label{doubletraceterm}
\\
&&
\hskip1cm
+ 2 \, \Big[ 
\gamma_{34;512} I^{(B)}  (34;512) 
+\gamma_{34;125} I^{(B)}  (34;125) 
+\gamma_{34;251} I^{(B)}  (34;251) 
\Big]
\nn\\
&&
\hskip1cm
+ 2\,\Big[ 
\gamma_{45;312} I^{(B)}  (45;312) 
+\gamma_{45;123} I^{(B)}  (45;123) 
+\gamma_{45;231} I^{(B)}  (45;231) 
\Big]
\nn\\
&&
\hskip1cm
+ 2\,\Big[ 
\gamma_{53;412} I^{(B)}  (53;412) 
+\gamma_{53;124} I^{(B)}  (53;124) 
+\gamma_{53;241} I^{(B)}  (53;241) 
\Big]
\bigg) \,.
\nn
\ea

Carrasco and Johansson derived a number of additional
relations satisfied by the numerator functions
$\beta_{12345}$  and $\gamma_{12:345}$.
In addition to the symmetries noted above,
$\gamma_{12;345}$ 
is invariant under $3 \leftrightarrow 4$,
and therefore under all permutations of $345$,
and so can be denoted simply as $\gamma_{12} = -\gamma_{21}$.
The box integral contribution to the double-trace amplitude 
(\ref{doubletraceterm})
can therefore be expressed as
\be
A_{5;3}(12;345) \big|_{\rm box}
=2 i g^5 \, \left[ \gamma_{34} I^{(B)}  (34) 
+ \gamma_{45} I^{(B)}  (45) 
+ \gamma_{53} I^{(B)}  (53) \right]
\ee
where $I^{(B)}(12)$ represents the sum of box integrals
\be
I^{(B)}(12)  \equiv 
I^{(B)}  (12;345) + I^{(B)}  (12;453) +I^{(B)}  (12;534) 
\label{truncbox}
\ee
which is invariant under $1 \leftrightarrow 2$ and
all permutations of 345.

In addition to the above symmetres, the  $\gamma_{ij}$ obey 
\be
\sum_{j=1}^5 \gamma_{ij} = 0
\label{sumgamma}
\ee
so that there are only six linearly-independent functions: 
$\gamma_{12}$, $\gamma_{13}$, $\gamma_{14}$, 
$\gamma_{23}$, $\gamma_{24}$, and $\gamma_{34}$.
The $\beta_{12345}$ 
are expressible in terms of the $\gamma_{ij}$:
\be
\beta_{12345} = {1 \over 2} 
\left( \gamma_{12}+ \gamma_{13}+ \gamma_{14}+ 
\gamma_{23}+ \gamma_{24}+ \gamma_{34}
\right)
\label{betasumgamma}
\ee
and vice versa
\be
\gamma_{12} = \beta_{12345}-\beta_{21345}
\label{betagamma}
\ee
which implies that the $\beta_{ijklm}$ satisfy
\be
\beta_{[ij][kl]m} =0.
\ee
In the next section, we will use these properties to
prove a one-loop relation between subleading-color SYM
and supergravity five-point amplitudes.

\section{One-loop supergravity-SYM relation}
\label{secsgvsym}
\setcounter{equation}{0}

Given the $\cN=4$ SYM amplitude in a form (\ref{cjfirst}) satisfying 
color-kinematic duality,
the BCJ conjecture \cite{Bern:2010ue} 
holds that the $\cN=8$ supergravity amplitude 
can be written as a double copy \cite{Carrasco:2011mn}
\be
\cM
= 
- \left(\kappa \over 2\right)^5
\sum_{S_5} 
\left( \frac{1}{10} 
\tbeta_{12345} \beta_{12345} I^{(P)} (12345) 
+
\frac{1}{4} 
\tgamma_{12;345}  \gamma_{12;345}  I^{(B)} (12;345) 
\right)
\label{bcjsupergravity}
\ee
where 
$\tbeta_{12345}$ and $\tgamma_{12;345}$
correspond to a second copy of the $\cN=4$ SYM numerators.
It was verified in refs.~\cite{Carrasco:2011mn,Bern:2011rj}
that \eqn{bcjsupergravity} is equivalent to the 
known five-point amplitude \cite{Bern:1998sv}.
The additional symmetries of $\gamma_{ij}$ and $\tgamma_{ij}$,
allow one to rewrite \eqn{bcjsupergravity} as 
\be
\cM
= 
- \left(\kappa \over 2\right)^5
\left(
\sum_{S_5/\Z_5 \times \Z_2} 
\tbeta_{12345} \beta_{12345} I^{(P)} (12345) 
+
\sum_{S_5/\Z_2 \times S_3} 
\tgamma_{12}  \gamma_{12}  I^{(B)} (12)
\right)
\label{supergravity}
\ee
where $I^{(B)}(12)$ is defined in \eqn{truncbox}.

The one-loop supergravity amplitude has a $1/\eps$ IR divergence
\cite{Dunbar:1995ed},
and so it is plausible that it may be expressed as a linear 
combination of the double-trace (subleading-in-$1/N$) color-ordered
amplitudes $A_{5;3}(12;345)$, which also have a $1/\eps$ leading
IR divergence.   We now establish such a relation. 

Consider the following linear combination of double-trace coefficients
\ba
\sum_{S_5} \tbeta_{12345}  A_{5;3} (12;345)
&=& 
\sum_{S_5} {1 \over 2} 
\left( \tbeta_{12345}  -\tbeta_{12435} \right) A_{5;3} (12;345)
\nn\\
&=&
\sum_{S_5} {1 \over 2} 
\left( \tbeta_{34512}  -\tbeta_{43512} \right) A_{5;3} (12;345)
\nn\\
&=&
\sum_{S_5} {1 \over 2} 
\tgamma_{34} A_{5;3} (12;345) 
\nn
\\
&=&
\sum_{S_5} {1 \over 6} 
(\tgamma_{34}+\tgamma_{45}+\tgamma_{53})  A_{5;3} (12;345) 
\ea
where we have used \eqn{betagamma} as well as 
the symmetries of $A_{5;3} (12;345)$ and $\beta_{12345}$.
Since
$\tgamma_{34}+\tgamma_{45}+\tgamma_{53}$
has the same symmetries as $A_{5;3} (12;345)$,
we can restrict the sum to the ten independent double-trace
coefficients
\be
\sum_{S_5} \tbeta_{12345}  A_{5;3} (12;345)
=
\sum_{S_5/\Z_2 \times S_3} 2
(\tgamma_{34}+\tgamma_{45}+\tgamma_{53})  A_{5;3} (12;345)  \,. 
\label{result}
\ee 
Each of the ten subleading-color amplitudes $A_{5:3}(12;345)$ appearing in
\eqn{result}
can be rewritten in terms of 
pentagon integrals $I^{(P)} (ijklm)$ and 
(sums of) box integrals $I^{(B)} (ij)$
using \eqn{doubletraceterm}.

First we extract the coefficient of $I^{(P)} (12345)$ in 
the double-trace amplitude $A_{5;3} (ij; klm)$.
These can be read off the first row of \eqn{transmatrix}.
Each of the following subleading-color amplitudes contain
a factor of $\beta_{12345}  I^{(P)} (12345)$:
\ba 
&&
A_{5:3} (12;345) , \quad
A_{5:3} (23;451) , \quad
A_{5:3} (34;512) , \quad
A_{5:3} (45;123) , \quad
A_{5:3} (51;234) ,
\nn\\
&&
A_{5:3} (13;245) , \quad
A_{5:3} (24;351) , \quad
A_{5:3} (35;412) , \quad
A_{5:3} (41;523) , \quad
A_{5:3} (52;134) .
\ea
Hence 
the coefficient of $i g^5 {I^{(P)}(12345)}  $
in $ \sum_{S_5} \tbeta_{12345}  A_{5;3} (12;345) $
is
\ba
&&
2 \Big[ (\tgamma_{34}+\tgamma_{45}+\tgamma_{53}) 
+ (\tgamma_{45}+\tgamma_{51}+\tgamma_{14}) 
+ (\tgamma_{51}+\tgamma_{12}+\tgamma_{25}) 
+ (\tgamma_{12}+\tgamma_{23}+\tgamma_{31})  
\nn\\ &&
+ (\tgamma_{23}+\tgamma_{34}+\tgamma_{42})
+ (\tgamma_{24}+\tgamma_{45}+\tgamma_{52}) 
+ (\tgamma_{35}+\tgamma_{51}+\tgamma_{13}) 
+ (\tgamma_{41}+\tgamma_{12}+\tgamma_{24}) 
\nn\\ &&
+ (\tgamma_{52}+\tgamma_{23}+\tgamma_{35}) 
+ (\tgamma_{13}+\tgamma_{34}+\tgamma_{41})
\Big]
\beta_{12345}
\nn\\
&=&
10 \,(
\tgamma_{12} + 
\tgamma_{13} + 
\tgamma_{14} + 
\tgamma_{23} + 
\tgamma_{24} + 
\tgamma_{34})
\beta_{12345}
\nn\\
&=& 
20 \,\tbeta_{12345} \beta_{12345}
\ea
where we used \eqns{sumgamma}{betasumgamma}
in the  last two lines.

Next we extract the coefficient of $I^{(B)}(34)$ in \eqn{result}.
The three independent subleading-color amplitudes that contribute
to the coefficient of $I^{(B)}(34)$ are 
\ba
A_{5;3}(12;345) \big|_{\rm box}
&=& 
2 \, \left[  \gamma_{34} I^{(B)}  (34) 
+ \gamma_{45} I^{(B)}  (45) 
+ \gamma_{53} I^{(B)}  (53) 
\right]
\nn\\
A_{5;3}(25;341) \big|_{\rm box}
&=& 
  2 \, \left[\gamma_{34} I^{(B)}  (34) 
+ \gamma_{41} I^{(B)}  (41) 
+ \gamma_{13} I^{(B)}  (13) 
\right]
\nn\\
A_{5;3}(51;342) \big|_{\rm box}
&=& 
  2 \, \left[\gamma_{34} I^{(B)}  (34) 
+ \gamma_{42} I^{(B)}  (42) 
+ \gamma_{23} I^{(B)}  (23) 
\right] \,.
\ea
Hence
the coefficient of $ig^5 I^{(B)}(34)$ 
in $ \sum_{S_5} \tbeta_{12345}  A_{5;3} (12;345) $
is
\ba
&&
 4 \left[ (\tgamma_{34} + \tgamma_{45} + \tgamma_{53} )
+ (\tgamma_{34} + \tgamma_{41} + \tgamma_{13} )
+ (\tgamma_{34} + \tgamma_{42} + \tgamma_{23} )
\right] \gamma_{34} 
\nn\\
&=&
 12 \, \tgamma_{34} \gamma_{34} 
 + 4\, (\tgamma_{45} + \tgamma_{41} + \tgamma_{42} )\gamma_{34} 
 + 4\, ( \tgamma_{53} + \tgamma_{13} + \tgamma_{23} )\gamma_{34} 
\nn\\
&=&  20 \, \tgamma_{34} \gamma_{34} 
\ea
where we used \eqn{sumgamma}
in the last line.

Assembling all the pieces we have
\be
\sum_{S_5} \tbeta_{12345}  A_{5;3} (12;345)
=
20 i g^5 \left(  
\sum_{S_5/\Z_5 \times \Z_2} \tbeta_{12345} \beta_{12345} I^{(P)} (12345)
+ 
\sum_{S_5/\Z_2 \times S_3} \tgamma_{12} \gamma_{12}  I^{(B)}(12)
\right) \,.
\ee
Comparing this with the one-loop supergravity amplitude (\ref{supergravity}),
we obtain the SYM-supergravity relation 
\be
\cM
= 
 {-1\over 20 i g^5} \left(\kappa \over 2\right)^5
\sum_{S_5} \tbeta_{12345}  A_{5;3} (12;345)
\label{generalization}
\ee
which is the five-point generalization of the one-loop
SYM-supergravity relation for the four-point 
amplitude \cite{Green:1982sw,Naculich:2008ys}.

We note that \eqn{generalization} is distinct from the 
relation between the one-loop supergravity amplitude 
and the {\it leading-color}  SYM amplitudes
\be
\cM
= 
 {-1\over i g^5} \left(\kappa \over 2\right)^5
\sum_{S_5/\Z_5 \times \Z_2}  
\tbeta_{12345}  A_{5;1} (12345)
\label{bbj}
\ee
proposed in eq.~(3.27) of ref.~\cite{Bern:2011rj}.
Indeed, substituting \eqn{comparing} into \eqn{bbj} 
one finds
\ba
\cM
&=& 
- \left(\kappa \over 2\right)^5
\sum_{S_5}  
{1 \over 10} \tbeta_{12345} 
\bigg(
\beta_{12345} I^{(P)} (12345) 
+
5 \gamma_{12;345}  I^{(B)} (12;345) 
\bigg) 
\nn\\
&=&
- \left(\kappa \over 2\right)^5
\sum_{S_5}  
\bigg[ {1 \over 10} \tbeta_{12345} 
\beta_{12345} I^{(P)} (12345) 
\nn\\
&&
\hskip2cm 
+ {1 \over 8} \left(\tbeta_{12345} - \tbeta_{21345} 
            +\tbeta_{12543} - \tbeta_{21543}  \right)
\gamma_{12;345}  I^{(B)} (12;345) 
\bigg]  \,.
\ea
Then by using the symmetries of $\tbeta_{12345}$ together with
the relation (\ref{betagamma}) one verifies that this is
precisely equal to \eqn{bcjsupergravity}, thus
providing a direct verification of the equivalence of 
eqs.~(3.27) and (3.28) of ref.~\cite{Bern:2011rj}.
Because the leading-color SYM amplitudes goes as $1/\eps^2$,
the leading IR divergences must cancel between the terms of 
\eqn{bbj}, whereas in \eqn{generalization} the individual
terms all go as $ 1 / \eps$.

\section{Conclusions}
\setcounter{equation}{0}

In this paper, we have recast the Carrasco-Johansson representation
of the one-loop $\cN=4$ SYM  five-point amplitude,
which respects color-kinematic duality, into a
trace basis consisting of leading- and subleading-color
partial amplitudes.
We then proposed and proved a linear relation
between the one-loop $\cN=8$ supergravity five-point
amplitude and the subleading-color SYM amplitude,
guided by the fact that the leading IR divergence
of each goes as $1/\eps$.
This generalizes earlier one- and two-loop relations
for four-point amplitudes \cite{Green:1982sw,Naculich:2008ys}.
In all of these cases, the kinematic numerator factors 
are loop-momentum-independent 
and so we were able to obtain relations
between integrated amplitudes, and not just integrands.
It is a challenging problem to find similar relations 
when numerator factors depend on loop momenta. 
Focusing on the IR behavior of the supergravity
and subleading-color SYM amplitudes may provide clues for
discovering further supergravity-SYM relations.

\section*{Acknowledgments}
We would like to thank H.~Nastase for his collaboration 
on our previous work on SYM-supergravity relations.

\appendix
\setcounter{equation}{0}

\section{IR divergence of the
subleading-color amplitude}

In this appendix, we rederive the result \cite{Nastase:2011mx}
that the leading IR divergence of the 
one-loop subleading-color five-point amplitude 
$A_{5;3} (12;345)$ goes as $1/\eps$. 
Recall from \eqn{Athirteen} that the one-loop
five-point coefficients satisfy
\be
  A^\One_{13} 
= A^\One_{1} -A^\One_{2} +A^\One_{3} -A^\One_{4} 
 -A^\One_{5} +A^\One_{6} -A^\One_{7} +A^\One_{8} 
 -A^\One_{9} +A^\One_{10} +A^\One_{11} -A^\One_{12}  \,.
\ee
Using this relation, it was shown in ref.~\cite{Nastase:2011mx} that 
the subleading-color amplitude $A_{5;3} (12;345)$ can be interpreted as
sums of volumes of simple polytopes \cite{ArkaniHamed:2010gg,Mason:2010pg}.

We know from ref.~\cite{Bern:1994zx} that the
one-loop planar five-point amplitude has a Laurent expansion beginning
\be
A^\One_i 
= - {5 g^2 N\over 32 \pi^2 \eps^2}  A^\Zero_i 
+ \cO\left( {1 \over \eps} \right),  \qquad i=1, \cdots 12
\ee
where $A^\Zero_i$ is the tree-level amplitude, 
so
\ba
A^\One_{13} 
&=& 
- {5 g^2 N\over 32 \pi^2 \eps^2} 
\Big( A^\Zero_{1} -A^\Zero_{2} +A^\Zero_{3} -A^\Zero_{4} 
-A^\Zero_{5} +A^\Zero_{6} \nn\\
 && 
\hskip13mm
 -A^\Zero_{7} +A^\Zero_{8} 
 -A^\Zero_{9} +A^\Zero_{10} +A^\Zero_{11} -A^\Zero_{12} 
\Big)
+ \cO\left( {1 \over \eps} \right) \,.
\label{Athirteenone}
\ea
The twelve tree-level coefficients
appearing in this equation are not independent, however,
but are related 
the Kleiss-Kuijf relations \cite{Kleiss:1988ne}.
A nice way to derive these relations 
is to write the tree-level amplitude
in a color basis consisting of \cite{DelDuca:1999ha,DelDuca:1999rs}  
\ba
&&
c_{1} = \f^{a_1 a_2 b}\f^{b a_3 c}\f^{c a_4 a_5}\,, \hskip 0.8cm  
c_{2} = \f^{a_1 a_4 b}\f^{b a_3 c}\f^{c a_2 a_5}\,, \hskip 0.8cm  
c_{3} = \f^{a_1 a_3 b}\f^{b a_4 c}\f^{c a_2 a_5}\,, \nn \\
&&
c_{4} = \f^{a_1 a_2 b}\f^{b a_4 c}\f^{c a_3 a_5}\,, \hskip 0.8cm  
c_{5} = \f^{a_1 a_4 b}\f^{b a_2 c}\f^{c a_3 a_5}\,, \hskip 0.8cm  
c_{6} = \f^{a_1 a_3 b}\f^{b a_2 c}\f^{c a_4 a_5}\,.
\ea
(These six basis elements are denoted 
$\{c_1, c_6, c_9, c_{12}, c_{14}, c_{15} \}$ in ref.~\cite{Bern:2008qj}.
The other nine $c_i$ are related to these by Jacobi identities.)
This color basis can be rewritten using a basis of
single-trace terms 
\be
c_i = \sum_{\la=1}^{12}  \M_{i\la} {\T_\la }
\ee
where $\T_\la$ are defined as in \eqn{singletrace}
except that the factor of $N$ is omitted at tree level.
Then
\be
\M_{i\la} 
= 
\left(
\begin{array}{cccccccccccc}
1 & 0 & 0 & 0 & 0 & 0~ & 1 & 0 & 0 & 0 & 1 & 1 \\
0 & 1 & 0 & 0 & 0 & 0~& 0 & 1 & 1 & 1 & 0 & 0 \\
0 & 0 & 1 & 0 & 0 & 0~& 1 & 1 & 1 & 0 & 0 & 0 \\
0 & 0 & 0 & 1 & 0 & 0~& 0 & 0 & 0 & 1 & 1 & 1 \\
0 & 0 & 0 & 0 & 1 & 0~& 0 & 0 & 1 & 1 & 1 & 0 \\
0 & 0 & 0 & 0 & 0 & 1~& 1 & 1 & 0 & 0 & 0 & 1 \\
\end{array}
\right) \,.
\ee
This rank-six matrix
possesses six null eigenvectors $R_{\la j}$:
\be
\sum_{\la = 1}^{12}   \M_{i\la} R_{\la j} = 0, \qquad  j = 1, \cdots 6 \,.
\ee
Given that $M_{i \la}$ has the form 
$\begin{pmatrix} \one_{6 \times 6}  & m \end{pmatrix}$,
one sees that 
$R = \begin{pmatrix} -m \\ \one_{6 \times 6} \end{pmatrix}$.
The existence of these null eigenvectors, 
together with \eqn{relatecoeff},
implies six relations among the color-ordered amplitudes
\be
\sum_{\la=1}^{12}  A^\Zero_\la R_{\la j} = 0, \qquad  j = 1, \cdots 6
\ee
which are precisely the Kleiss-Kuijf relations \cite{Kleiss:1988ne}
at five points.
These relations can be used to eliminate 
$A_7^\Zero$ through $A_{12}^\Zero$, 
\eg,  $ A_7^\Zero = A_1^\Zero+ A_3^\Zero+ A_6^\Zero $.
Substituting these relations into \eqn{Athirteenone}
demonstrates that the leading $1/\eps^2$ IR divergence of 
$A_{5;3} (12;345)$ vanishes.

\vfil\break

\end{document}